\documentclass[aps,pra,reprint,longbibliography]{revtex4-2}

\usepackage{graphicx}% Include figure files
\usepackage{dcolumn}% Align table columns on decimal point
\usepackage{bm}% bold math
\usepackage{amsmath,amssymb,amsthm}
\usepackage{physics}          % \bra, \ket, \grad, etc.
\usepackage{mathtools}
\usepackage{hyperref}
\usepackage{cleveref}
\usepackage{xcolor}
\usepackage{booktabs}

\renewcommand{\hbar}{\hslash}

\newcommand{\kap}{\kappa}
\newcommand{\LS}{Lohmiller-Slotine}

\theoremstyle{plain}

\begin{document}
\title{Quantum tunneling, global phases and the limits of classical action reconstructions}

\author{Chong Qi} \email{chongq@kth.se}

 \affiliation{Department of Physics, KTH Royal Institute of Technology, AlbaNova University Center, S-106 91, Stockholm, Sweden}
\author{Mário B. Amaro}
 \affiliation{Department of Physics, Stockholm University, AlbaNova University Center, S-106 91, Stockholm, Sweden}
\date{\today}

\begin{abstract}
%It was proposed recently that the Schr\"{o}dinger wave function can be reconstructed exactly from a discrete superposition of classical action branches weighted by associated classical densities, without invoking semiclassical approximations. We examine this construction for quantum tunneling through finite potential barriers and for quantum phase phenomena. While the construction is formally consistent with the decomposition of the wave function for systems for which the Hamilton-Jacobi equation admits globally defined real branches, we show that it breaks down in classically forbidden regions where no real classical action exists. Using the rectangular barrier and Coulomb-barrier tunneling in alpha decay and nuclear fusion as explicit examples, we demonstrate that the wave function necessarily requires either a non-vanishing quantum potential or complex-valued action. In the finite-barrier problem, the exponentially growing component inside the barrier, fixed by global boundary conditions, is essential for transmission and cannot be generated from local real classical trajectories alone. We further show that Berry phase, flux quantization, Josephson tunneling, and dc SQUID interference introduce global quantum phase constraints that cannot be reproduced from local classical action transport unless the relevant quantum phase structure is inserted a posteriori. The analysis clarifies the intrinsically non-classical character of tunneling and phase-coherent quantum phenomena, and the restricted domain of applicability of classical-action reconstructions.
It was proposed recently that the Schr\"odinger wave function can be reconstructed exactly from a discrete superposition of classical action branches weighted by associated classical densities, without semiclassical approximations. We examine this construction for quantum tunneling through finite potential barriers and for quantum phase phenomena. Although formally consistent when the Hamilton-Jacobi equation admits globally defined real branches, the construction breaks down in classically forbidden regions where no real classical action exists. Using rectangular and Coulomb barrier tunneling in alpha decay and nuclear fusion, we show that the wave function requires either a non-vanishing quantum potential or complex-valued action. The growing barrier component fixed by global boundary conditions is essential for transmission and cannot arise from local real classical trajectories alone. Berry phase, flux quantization, Josephson tunneling, and dc SQUID interference likewise impose global phase constraints absent from local classical action transport.
\end{abstract}

\maketitle

\section{Introduction}
\label{sec:intro}
The relationship between classical and quantum mechanics remains one of the central conceptual problems of theoretical physics. Since the early development of quantum theory, considerable effort has been devoted to understanding how quantum dynamics emerges from, or relates to, underlying classical structures. These include Bohr’s correspondence principle, Hamilton–Jacobi theory, semiclassical approximations, and Feynman’s path-integral formulation \cite{Bohr1920,Feynman1948}. Despite these developments and many subsequent advances, a fully satisfactory reconstruction of quantum mechanics from purely classical dynamical principles remains elusive.
This long-standing problem is revisited in Ref.~\cite{LS2026}, which claims that the Schr\"odinger equation can be solved exactly using only classical action branches and associated classical densities. In that formulation, the quantum wave function is expressed as
\begin{equation}
\psi = \sum_{j}\sqrt{\rho_j}e^{i\phi_j/\hbar},
\label{eq:LS_ansatz}
\end{equation}
where $\rho_j$ denotes a classical probability density transported along a specific branch $j$ via the classical continuity equation, and the classical-action phase ${\phi_j}$ satisfies the classical Hamilton–Jacobi (HJ) equation.

The class of systems to which Ref.~\cite{LS2026} applies is, however, implicitly restricted to those admitting globally defined, real, multi-valued solutions of the HJ equation—a highly non-generic condition. Their treatment of tunneling obscures a critical physical and mathematical breakdown. Although presented as a ``barrier problem”, the example is in fact a semi-infinite potential step rather than a finite barrier through which transmission can occur.

We are therefore interested in extending the classical-action formalism to the physically essential case of quantum tunneling through a finite potential barrier, arguably the paradigmatic quantum effect with no classical analogue. It is widely regarded as a cornerstone of modern quantum physics, underpinning phenomena ranging from nuclear decay to electronic transport and  its conceptual depth continues to drive both theoretical and experimental developments \cite{Bender2011,Anastopoulos2017,Turok2014,Bramberger2016,PhysRevD.111.085027,nishimura2023newpicture}.  A formulation deriving tunneling purely from real classical action would therefore be remarkable, but faces a fundamental obstacle: tunneling occurs precisely in classically forbidden regions, where no real classical action exists, raising the question of whether tunneling can be reproduced within a framework constructed solely from real classical action.

%Our analysis shows that, while the decomposition proposed in Ref.~\cite{LS2026} remains formally consistent for systems admitting globally defined real branches, it breaks down for finite-barrier tunneling. For both examples of rectangular barrier and Coulomb-barrier tunneling in alpha decay, the exact wave function necessarily requires either a non-vanishing quantum potential or complex-valued action in the forbidden region.

Our analysis shows that, while the decomposition proposed in Ref.~\cite{LS2026} remains formally consistent for systems admitting globally defined real branches, it breaks down in classically forbidden regions and in quantum phenomena governed by global phase constraints. For rectangular-barrier tunneling, Coulomb barrier tunneling in alpha decay, and Coulomb  barrier penetration in nuclear fusion, the exact wave function necessarily requires either a non-vanishing quantum potential or complex-valued action. We further show that Berry phase, flux quantization, Josephson tunneling, and dc SQUID interference cannot be reconstructed from local classical action transport alone, since their description depends on geometric phase, macroscopic phase single-valuedness, or coherent tunneling amplitudes fixed by global boundary conditions.

\subsection{Historical background and the classical action}
\label{sec:history}
In the same year as the introduction of the Schr\"odinger equation, Madelung~\cite{Madelung1926}
showed that, by assuming that the corresponding wave function can be decomposed in the polar form 
\begin{align}\Psi = \sqrt{\rho}\,e^{iS/\hbar},\end{align}
 the  equation can be transformed into two coupled real equations:
\begin{align}
  \frac{\partial \rho}{\partial t} + \nabla \cdot \left(\rho \,\frac{\nabla S}{m}\right) &= 0 ,
  \label{eq:continuity}\\
  \frac{\partial S}{\partial t} + \frac{(\nabla S)^2}{2m} + V + Q &= 0 ,
  \label{eq:QHJ}
\end{align}
where $\rho(x,t) > 0$ and
$S(x,t)$ are real quantities and
\begin{equation}
  Q(\mathbf{x},t) = -\frac{\hbar^2}{2m}\frac{\nabla^2 \sqrt{\rho}}{\sqrt{\rho}}
  \label{eq:Qpot}
\end{equation}
is the so-called \emph{quantum potential}.

Eq.~\eqref{eq:continuity} is the classical continuity equation, while Eq.~\eqref{eq:QHJ} resembles the classical HJ equation but contains the extra term $Q$. If $Q$ were zero, Eq.~\eqref{eq:QHJ} would reduce to classical mechanics exactly. The quantum potential is \emph{not} a small correction in the semiclassical sense: it is $\mathcal{O}(\hbar^2)$ at leading order but can become dominant in regions where $\rho$ varies rapidly or approaches zero, particularly near turning points and inside classically forbidden regions.

de~Broglie~\cite{deBroglie1927} interpreted $S$ as a real
guiding phase and $\nabla S/m$ as a particle velocity, giving a
deterministic trajectory picture.  Bohm~\cite{Bohm1952a,Bohm1952b}
subsequently developed this into a self-consistent pilot-wave formulation in which $Q$ acts as a genuine potential governing
particle trajectories. 

Closely related ideas also appear in semiclassical theory. Van~Vleck~\cite{VanVleck1928} derived the leading semiclassical
approximation to the quantum propagator in terms of classical paths,
already recognizing the need to sum over \emph{multiple} classical
trajectories. Ref. \cite{LS2026} may be seen as a non-perturbative
generalization of the Van~Vleck propagator to situations where the
multi-valuedness is exact (not semiclassical), which is indeed a step
forward for the class of problems where real branches exist globally. A similar program was outlined by Tipler
\cite{Tipler2010}, who noted that Schr\"{o}dinger's equation can be
reduced to HJ solutions in special cases.
In addition, the algebraic approach of
Koopman~\cite{Koopman1931} and more recent works~\cite{Morgan2020}
provide an operator-algebraic framework in which classical mechanics
is embedded in a Hilbert space, which may make it formally analogous to quantum
mechanics.

For systems admitting globally defined real solutions of the HJ equation, the classical-action formulation of Ref. \cite{LS2026} is formally consistent with the Madelung decomposition.
 For any
solution $\psi(x,t)$ of the Schr\"{o}dinger equation
\begin{equation}
  i\hbar\frac{\partial\psi}{\partial t}
  = -\frac{\hbar^2}{2m}\nabla^2\psi + V(x)\psi,
  \label{eq:SE}
\end{equation}
direct substitution of $\psi_j = \sqrt{\rho_j}\,e^{i\phi_j/\hbar}$
into~\eqref{eq:SE} to the left-hand side gives:
\[
  i\hbar\partial_t\psi_j
  = \left(i\hbar\frac{\partial_t\sqrt{\rho_j}}{\sqrt{\rho_j}}
         - \partial_t\phi_j\right)\psi_j.
\]
Computing the Laplacian:
\begin{equation*}
\begin{split}
\nabla^2\!\left(\sqrt{\rho_j}\,e^{i\phi_j/\hbar}\right)
  =& \left[\frac{\nabla^2\!\sqrt{\rho_j}}{\sqrt{\rho_j}}
    + \frac{2i}{\hbar}\frac{\nabla\sqrt{\rho_j}}{\sqrt{\rho_j}}
      \cdot\nabla\phi_j\right.\\
    &\quad\quad\quad\left. - \frac{(\nabla\phi_j)^2}{\hbar^2}
    + \frac{i}{\hbar}\nabla^2\phi_j\right]\psi_j,
\end{split}
\end{equation*}
and substituting into the right-hand side ~\eqref{eq:SE} and dividing the equation by $\psi_j \neq 0$ gives:
\begin{equation*}
\begin{split}
  i\hbar\frac{\partial_t\sqrt{\rho_j}}{\sqrt{\rho_j}} - \partial_t\phi_j
  &= -\frac{\hbar^2}{2m}\frac{\nabla^2\!\sqrt{\rho_j}}{\sqrt{\rho_j}} + \frac{(\nabla\phi_j)^2}{2m} + V
    \\
  &\quad\quad - \frac{i\hbar}{2m}\frac{2\nabla\sqrt{\rho_j}\cdot\nabla\phi_j
      + \rho_j\nabla^2\phi_j}{\rho_j}.
\end{split}
\end{equation*}
The imaginary part gives
exactly the continuity equation~\eqref{eq:continuity} for $\rho_j$; the
real part gives~\eqref{eq:QHJ} which differs from the classical HJ equation 
by the quantum potential $Q$. 

The 
claim that each individual $\phi_j$ satisfies the classical HJ equation is only
consistent with~\eqref{eq:SE} when $Q_j = 0$. 
To satisfy the Schr\"{o}dinger equation without $Q$,
one requires either:
\begin{enumerate}
  \item[(a)] $Q_j = 0$ for each branch individually , or
  \item[(b)] The quantum potentials from different branches
    cancel exactly in the superposition.
\end{enumerate}
Condition~(b) indeed occurs in several exactly solvable systems. However, in such cases, the reconstruction is not generated purely from classical inputs: the relative phases and amplitudes between branches encode intrinsically quantum information through highly constrained interference conditions.

The condition $Q_j=0$ for each branch is itself extremely restrictive. Since
\begin{equation}
  \nabla^2\!\sqrt{\rho_j} = 0,
  \label{eq:harmonic}
\end{equation}
the amplitude $\sqrt{\rho_j}$ must be a \emph{harmonic function}. The general solutions are linear combinations of harmonic polynomials
and, in one dimension, simply $\sqrt{\rho_j} = ax + b$.  
This immediately excludes any potential with spatially varying density, which is essentially all physically interesting cases.
For a
probability density, $\rho_j \geq 0$, so $\sqrt{\rho_j}$ must be
non-negative, further restricting the class.
 
The essential question is therefore not whether the decomposition
can formally reproduce solutions of the Schr\"odinger equation, but whether the individual branches can consistently retain the interpretation of real classical actions in physically nontrivial situations, particularly in classically forbidden regions.

 \section{Potential step and the need for complex action}
\label{sec:step}
Consider the standard one-dimensional potential step problem \cite{Griffiths2018,Flugge1999,Qi2026}, also studied in Ref. \cite{LS2026}: a particle of mass $m$ and energy $E > 0$ incident from the
left on the potential step
\begin{equation}
  V(x) = \begin{cases} 0 & x < 0, \\ V_0 > 0 & x \geq 0. \end{cases}
  \label{eq:V_step}
\end{equation}
We treat the sub-barrier case $E < V_0$.  The exact stationary
solution of~\eqref{eq:SE} is:
\begin{align}
  \psi_{\mathrm{I}}(x,t)
    &= \Bigl(e^{ikx} + r\,e^{-ikx}\Bigr)e^{-iEt/\hbar},
    \quad x < 0,
    \label{eq:psi_I}\\
  \psi_{\mathrm{II}}(x,t)
    &= A\,e^{-\kap x}\,e^{-iEt/\hbar},
    \quad x \geq 0,
    \label{eq:psi_II}
\end{align}
where
\begin{equation}
  k = \frac{\sqrt{2mE}}{\hbar} > 0, \qquad
  \kap = \frac{\sqrt{2m(V_0-E)}}{\hbar} > 0.
  \label{eq:k_kappa}
\end{equation}
Normalizability requires discarding the growing
exponential $e^{\kap x}$ since the domain extends to
$x \to +\infty$.  Matching $\psi$ and $\partial_x\psi$ at $x = 0$ gives:
\begin{equation}
  r = \frac{k - i\kap}{k + i\kap}, \qquad
  A = \frac{2k}{k + i\kap},
  \label{eq:r_A}
\end{equation}
with $|r|^2 = 1$ (total reflection), indicating zero transmission.
 
For the Madelung decomposition at $x \geq 0$, since $\psi_{\mathrm{II}} = \sqrt{\rho}\,e^{iS/\hbar}$ and
$\psi_{\mathrm{II}} = Ae^{-\kap x}e^{-iEt/\hbar}$ with $A$ complex, one can write $A = |A|e^{i\theta}$ with
$\theta = \arg A$:
\begin{equation}
  \sqrt{\rho} = |A|e^{-\kap x}, \qquad
  S = \hbar\theta - Et,
  \label{eq:polar_II}
\end{equation}
where $\theta = \arctan(-\kap/k)$ is a \emph{constant} (independent of
$x$).  Therefore:
\begin{equation}
  \partial_x S = 0, \qquad
  v = \frac{\partial_x S}{M} = 0.
  \label{eq:v_step}
\end{equation}
The equation on the right-hand side defines the Bohmian velocity, which is zero inside the potential step. 

The Bohmian velocity is a postulate (the guidance equation 
$v = \partial_x S / M$), not derived from the Schr\"odinger equation. 
What can be derived from the Schr\"odinger equation is the probability current
$j = \hbar\,\mathrm{Im}(\psi^* \partial_x \psi) / M$. 
One can define a velocity ($v_j = \partial_x \phi_j / M$ per branch) for the $\mathrm{LS}$ framework.
 
The quantum potential for $x \geq 0$ is
\begin{equation}
  Q_{\mathrm{step}} = -\frac{\hbar^2\kap^2}{2m}
  = -\frac{\hbar^2}{2m}\cdot\frac{2m(V_0-E)}{\hbar^2}
  = -(V_0 - E),
  \label{eq:Q_step}
\end{equation}
which gives an effective potential 
\begin{equation}
  V_{\mathrm{eff}} = V_0 + Q_{\mathrm{step}} = V_0 - (V_0 - E) = E.
  \label{eq:Veff_step}
\end{equation}
That is consistent with the expectation that the velocity is zero.

For completeness,
by writing $r = |r|e^{-2i\delta}$ with $\delta = \arctan(\kap/k)$
and $|r|=1$, one obtains the Madelung decomposition of the wave function for region I as
\begin{align}
  \sqrt{\rho_{\mathrm{I}}}
    &= \bigl|e^{ikx} + r\,e^{-ikx}\bigr|
     = 2|\cos(kx + \delta)|,
  \label{eq:rhoI_Bohm}\\
  S_{\mathrm{I}}(x,t)
    &= \hbar\,\arg\!\bigl(e^{ikx} + r\,e^{-ikx}\bigr) - Et\\
     &= \hbar\arctan\!\left(\frac{\sin(kx+\delta)}{\cos(kx+\delta)}\right) - Et
     \notag\\
    &= \hbar(kx + \delta) - Et.
  \label{eq:SI_Bohm}
\end{align}

 In the \LS{} (LS) multi-valued classical-action framework, there exist two real branches  for region I ($x < 0$), corresponding to incident ($+$) and
reflected ($-$) momenta $p = \pm\hbar k$:
\begin{eqnarray}
  \phi_+^{\mathrm{I}} =& \pm\hbar k x - Et,\\
  \phi_-^{\mathrm{I}} = &\pm\hbar k x - Et+\hbar\arg r.
  \label{eq:phiI_LS}
\end{eqnarray}
The densities are $\sqrt{\rho_+} = 1$ (incident) and
$\sqrt{\rho_-} = |r| = 1$ (reflected), both constant.
The wave function in region~I is:
\begin{equation}
\begin{split}
  \psi_{\mathrm{I}}
  &= \sqrt{\rho_+}\,e^{i\phi_+^{\mathrm{I}}/\hbar}
  + \sqrt{\rho_-}\,e^{i\phi_-^{\mathrm{I}}/\hbar} =\\
  &=\bigl(e^{ikx} + r\,e^{-ikx}\bigr)e^{-iEt/\hbar}.
  \label{eq:psiI_LS}
\end{split}
\end{equation}

Region II ($x \geq 0$), however, becomes problematic.
The classical HJ equation requires
$$(\partial_x\phi)^2 = 2m(E-V_0) < 0,$$
which has no real
solution.  Ref.~\cite{LS2026} therefore  introduces two branches with \emph{imaginary}
momenta $p_T = \pm i\hbar\kap$:
\begin{equation}
  \phi_\pm^{\mathrm{II}} = \pm i\hbar\kap x - Et.
  \label{eq:phiII_LS}
\end{equation}
The corresponding wave contributions are:
\begin{eqnarray}
      \sqrt{\rho_+}\,e^{i\phi_+^{\mathrm{II}}/\hbar}
  = \sqrt{\rho_+}\,e^{-\kap x}e^{-iEt/\hbar},\nonumber
  \\
  \sqrt{\rho_-}\,e^{i\phi_-^{\mathrm{II}}/\hbar}
  = \sqrt{\rho_-}\,e^{+\kap x}e^{-iEt/\hbar}.
  \label{eq:psiII_LS_both}
\end{eqnarray}
The growing term $e^{+\kap x}$ diverges as $x\to+\infty$
and is discarded by normalisability, leaving:
\begin{equation}
  \psi_{\mathrm{II}}^{\mathrm{LS}}
  = \sqrt{\rho_T}\,e^{-\kap x}e^{-iEt/\hbar}
  = \psi_{\mathrm{II}},
  \label{eq:psiII_LS}
\end{equation}
where $\sqrt{\rho_T} = |A|$. 
Already at this stage, one may say that the \LS{} framework ceases to have a classical action interpretation, because no real classical action exists in the forbidden region. The action $\phi_\pm = -Et \pm i\hbar\kap x$ is purely imaginary in its
spatial part. 

The Bohmian result of zero velocity
follows from the observation that $\psi_{\mathrm{II}}$ is
real up to a global constant phase, giving $\partial_x S = 0$
directly. No imaginary quantity is needed. On the other hand, it is claimed in Ref. \cite{LS2026}
that since the trajectory equation is linear for $x > 0$, superposition of the two imaginary-momentum solutions gives a real trajectory:
\begin{equation}
 m\dot{x} = p_+ + p_- = i\hbar\kap + (-i\hbar\kap) = 0,
  \label{eq:LS_v_claim}
\end{equation}
which they interpret as $\dot{x} = 0$.
However, this is incorrect because the $e^{+\kap x}$ term diverges as $x \to +\infty$ and must be discarded by normalizability. 

In addition, the HJ equation is nonlinear
due to the term $(\partial_x\phi)^2$: if $\phi_1$ and $\phi_2$
are solutions, $\phi_1 + \phi_2$ is generally not a solution,
since
\begin{equation*}
\begin{split}
  \frac{(\partial_x(\phi_1+\phi_2))^2}{2m}
  = \frac{(\partial_x\phi_1)^2 + 2\partial_x\phi_1\cdot\partial_x\phi_2
          + (\partial_x\phi_2)^2}{2m}\\
  \neq \frac{(\partial_x\phi_1)^2}{2m} + \frac{(\partial_x\phi_2)^2}{2m}.
\end{split}
\end{equation*}
The cross term $\partial_x\phi_1\cdot\partial_x\phi_2/m$ is
the obstruction.

\section{The Rectangular Barrier}
\label{sec:barrier}
True transmission does not occur in the sub-barrier potential-step problem. Instead, let us consider the rectangular potential barrier \cite{Qi2026}
\begin{equation}
  V(x) = \begin{cases}
    0   & x < 0, \\
    V_0 & 0 \leq x \leq a, \\
    0   & x > a,
  \end{cases}
  \label{eq:V_barrier}
\end{equation}
with $E < V_0$ and $0 < a < \infty$. The exact stationary solution is:
\begin{align}
  \psi_{\mathrm{I}}(x,t)
    &= \bigl(e^{ikx} + r\,e^{-ikx}\bigr)e^{-iEt/\hbar},
    \quad x < 0,
    \label{eq:psi_I_bar}\\
  \psi_{\mathrm{II}}(x,t)
    &= \bigl(Ae^{-\kap x} + Be^{+\kap x}\bigr)e^{-iEt/\hbar},
    \quad 0 \leq x \leq a,
    \label{eq:psi_II_bar}\\
  \psi_{\mathrm{III}}(x,t)
    &= t\,e^{ikx}e^{-iEt/\hbar},
    \quad x > a,
    \label{eq:psi_III_bar}
\end{align}
with $k$ and $\kap$ as in~\eqref{eq:k_kappa}.
 
Matching $\psi$ and $\partial_x\psi$ at $x = 0$ and $x = a$ gives
four equations for four unknowns $\{A, B, r, t\}$.  Matching at
$x = a$:
\begin{align}
  Ae^{-\kap a} + Be^{+\kap a} &= te^{ika},
  \label{eq:match_a1}\\
  -\kap Ae^{-\kap a} + \kap Be^{+\kap a} &= ikt\,e^{ika}.
  \label{eq:match_a2}
\end{align}
Adding and subtracting:
\begin{align}
  Be^{+\kap a} &= \frac{1}{2}\!\left(1 + \frac{ik}{\kap}\right)t\,e^{ika},
  \label{eq:B_exact}\\
  Ae^{-\kap a} &= \frac{1}{2}\!\left(1 - \frac{ik}{\kap}\right)t\,e^{ika}.
  \label{eq:A_exact}
\end{align}
Therefore:
\begin{equation}
  B = \frac{t}{2}\!\left(1 + \frac{ik}{\kap}\right)e^{ika-\kap a} \neq 0,
  \label{eq:B_nonzero}
\end{equation}
and the ratio:
\begin{equation}
  \frac{B}{A} = \frac{1 + ik/\kap}{1 - ik/\kap}\,e^{-2\kap a}.
  \label{eq:BA_ratio}
\end{equation}
The transmission coefficient is:
\begin{equation}
  \mathcal{T} = |t|^2
  = \left[1 + \frac{(k^2+\kap^2)^2}{4k^2\kap^2}\sinh^2(\kap a)\right]^{-1}.
  \label{eq:T_exact}
\end{equation}
In the thick-barrier limit $\kap a \gg 1$:
\begin{equation}
  \mathcal{T} \approx \frac{16k^2\kap^2}{(k^2+\kap^2)^2}\,e^{-2\kap a}.
  \label{eq:T_approx}
\end{equation}

The growing exponential $Be^{+\kap x}$ in~\eqref{eq:psi_II_bar} is
nonzero for any finite $a > 0$ and $\mathcal{T} > 0$.  Moreover,
$|B/A| = e^{-2\kap a}(1 + k^2/\kap^2)^{1/2}/(1+k^2/\kap^2)^{1/2}$
so $|B| < |A|$ with $|A|$ exponentially larger than $|t|$, and $|B|$ exponentially smaller than $|t|$:
\begin{equation}
  |A| = \frac{|t|}{2}\sqrt{1+\frac{k^2}{\kap^2}}\,e^{+\kap a},
  \qquad
  |B| = \frac{|t|}{2}\sqrt{1+\frac{k^2}{\kap^2}}\,e^{-\kap a}.
  \label{eq:AB_magnitudes}
\end{equation}
%The near-cancellation of these two exponentially large terms produces
%the exponentially small probability density  inside the barrier.
Since for a thick barrier $|t|\sim Ce^{-\kappa a}$, the finite-barrier is hence described as a dominant decaying component plus a small growing component whose coefficient is fixed by the exit boundary condition, effectively producing an exponentially small probability density inside the barrier.
 
\subsection{Madelung decomposition inside the barrier}
 
Let us define $\psi_{\mathrm{II}} = \sqrt{\rho_{\mathrm{II}}}\,e^{iS_{\mathrm{II}}/\hbar}$ and,
since $A$ and $B$ are both complex,
\begin{align}
  A &= |A|e^{i\alpha}, \qquad
  B = |B|e^{i\beta},
  \label{eq:AB_phases}
\end{align}
where 
\begin{align}
  \alpha &= ka + \arg\!\left(1 - \frac{ik}{\kap}\right)
          = ka - \arctan\!\frac{k}{\kap},
  \label{eq:alpha}\\
  \beta  &= ka + \arg\!\left(1 + \frac{ik}{\kap}\right)
          = ka + \arctan\!\frac{k}{\kap},
  \label{eq:beta}
\end{align}
the density and phase of $\psi_{\mathrm{II}}$ are:
\begin{align}
  \rho_{\mathrm{II}}(x)
    &= \bigl|Ae^{-\kap x} + Be^{+\kap x}\bigr|^2 \notag\\
    &= |A|^2e^{-2\kap x} + |B|^2e^{+2\kap x} +\\
       & + 2|A||B|\cos(\alpha - \beta),
  \label{eq:rho_II}\\[4pt]
  S_{\mathrm{II}}(x,t)
    &= \hbar\arg\!\left(|A|e^{i\alpha-\kap x}
       + |B|e^{i\beta+\kap x}\right) - Et.
  \label{eq:S_II}
\end{align}
More importantly, $S_{\mathrm{II}}(x,t)$ depends
nontrivially on $x$, giving a nonzero Bohmian velocity:
\[
  v_{\mathrm{II}}(x) = \frac{\partial_x S_{\mathrm{II}}}{m}
  = \frac{\hbar}{m}\,\partial_x\arg\!\left(|A|e^{i\alpha-\kap x}
    + |B|e^{i\beta+\kap x}\right) \neq 0.
  \label{eq:v_II}
\]
This shows that the Bohmian particle has a nonzero velocity inside a finite barrier. This contrasts sharply with the potential step (section~\ref{sec:step}), where $v = 0$. The physical reason is clear: for the step, only the decaying wave $e^{-\kap x}$ is present, giving a real wave function (up to overall phase) with constant spatial phase.  For the finite barrier, the growing wave $Be^{+\kap x}$ — forced by the exit boundary condition at $x = a$ — introduces a nontrivial spatial phase through the interference $Ae^{-\kap x} + Be^{+\kap x}$ with complex $A,B$. The quantum potential inside the barrier is
$$\begin{aligned}
Q_{\mathrm{II}}(x) = -\frac{\hbar^2}{2m} \cdot \frac{1}{2\rho_{\mathrm{II}}^2} \Big[ &4\kappa^2(|A|^2e^{-2\kappa x}+|B|^2e^{2\kappa x})\rho_{\mathrm{II}} \\
&- 2\kappa^2(|A|^2e^{-2\kappa x}-|B|^2e^{2\kappa x})^2 \Big]
\end{aligned}$$
The quantum potential is spatially varying, depends on $x$ through
$e^{\pm 2\kap x}$, and depends on $a$ through the coefficients
$|A|$ and $|B|$ given in~\eqref{eq:AB_magnitudes}.  It reduces to
$-(V_0-E)$ (the step value) only in the limit $a \to \infty$, $B \to 0$.

 For the {finite barrier}, the exit wall at $x = a$ forces the
growing exponential $Be^{+\kap x}$ to be nonzero.  This introduces a
nontrivial spatial phase gradient inside the barrier, giving a
nonzero Bohmian velocity.  The particle \emph{moves} through the
barrier, guided by the pilot wave.  The transmitted amplitude
$t \propto e^{-\kap a}$ is exponentially small, but it is nonzero:
a fraction $\mathcal{T} = |t|^2$ of the probability flux crosses.
This is genuine quantum tunnelling.

 \subsection{Nonlocal structure of the tunneling solution and limits of local classical branch reconstruction}

 In region~II ($0 \leq x \leq a$), the LS ansatz
$\psi_{\mathrm{II}} = \sqrt{\rho_+}\,e^{i\phi_+/\hbar}
+ \sqrt{\rho_-}\,e^{i\phi_-/\hbar}$ matches the exact
solution with complex actions
\begin{equation}
  \phi_\pm = \pm i\hbar\kap x - Et,
  \label{eq:phi_II}
\end{equation}
and complex densities
\begin{align}
  \sqrt{\rho_+} &= A
    = \frac{t}{2}\!\left(1 - \frac{ik}{\kap}\right)
      e^{ika+\kap a},
  \label{eq:rhoplus}\\
  \sqrt{\rho_-} &= B
    = \frac{t}{2}\!\left(1 + \frac{ik}{\kap}\right)
      e^{ika-\kap a},
  \label{eq:rhominus}
\end{align}

While the LS framework correctly predicts that the classical densities $\rho_\pm$ are spatially constant inside the square barrier (due to $\Delta_M \phi = 0$), these classical coefficients are complex and therefore encode global phase information fixed simultaneously by both boundaries and cannot be interpreted as classical probability densities. In standard wave mechanics, the values of $\rho_+$ and $\rho_-$ must be derived from the classical continuity equation at the branch point $x=0$. Because this is evaluated purely at $x=0$, it contains zero mathematical information regarding the barrier exit at $x=a$. It is mathematically impossible to solve a local algebraic flux equation at $x=0$ and yield constants $\rho_\pm$ with the global exponential dependence $e^{\pm 2\kap a}$. 

The fundamental difference is therefore: \emph{the growing
exponential $Be^{+\kap x}$, absent for the step and nonzero for the
barrier, is the carrier of tunnelling}.  Its amplitude is set by
the global boundary condition at the exit wall, encoding information
about the barrier width and shape.  No local, stationary construction can determine it from local
initial data.
In other words, a local classical action framework cannot mathematically generate the correct constant amplitudes required to construct a tunneling wave function. 

The rectangular barrier reduces to the step as $a \to \infty$.
From~\eqref{eq:B_nonzero}, $B \propto e^{-\kap a} \to 0$, and
from~\eqref{eq:T_approx}, $\mathcal{T} \to 0$.  In this limit:
\begin{itemize}
  \item $\psi_{\mathrm{II}} \to Ae^{-\kap x}e^{-iEt/\hbar}$ (real up to phase),
  \item $v_{\mathrm{II}} \to 0$,
  \item $Q_{\mathrm{II}} \to -(V_0-E)$,
  \item $\mathcal{T} \to 0$.
\end{itemize}
The LS construction correctly recovers the step result in this limit.
The LS treatment of tunneling is therefore the $a \to \infty$ limit
of the full barrier problem: it captures the evanescent tail, but
misses the tunnelling amplitude $e^{-\kap a}$ that is the entire
physical content of quantum tunnelling.

\section{Bound and Unbound States of Coulomb Potential}
\label{sec:coulomb}
Alpha decay provides the historically most important application of
quantum tunneling, with precisely measurable decay half-lives spanning
more than 20 orders of magnitude.  Gamow~\cite{Gamow1928} and
Gurney-Condon~\cite{GurneyCondon1928} independently applied 
tunneling to nuclear barrier penetration in 1928, providing the first
quantitative quantum calculation in nuclear physics.  We use this
example to make the failure of the LS construction explicit and
numerically concrete.

For a particle moving in a $1/r$ Coulomb field, the radial part of the Schr\"odinger equation can be written as:
\begin{equation}
    \left[ \frac{d^2}{d\varrho^2} + 1 - \frac{2\eta}{\varrho} - \frac{L(L+1)}{\varrho^2} \right] u_L(\varrho) = 0
    \label{eq:schrodinger_reduced}
\end{equation}
where $\varrho = kr$ (with $k = \frac{\sqrt{2mE}}{\hbar} $ being the wave number and $r$ the radial distance). $L$ is the orbital angular momentum quantum number. $\eta$ is the Sommerfeld parameter (or Coulomb parameter); $\eta = \frac{Z_1 Z_2 e^2}{\hbar} \sqrt{\frac{m}{2E}} > 0$ for a repulsive Coulomb barrier. The exact solutions to this differential equation are the regular ($F_L$) and irregular ($G_L$) Coulomb wave functions. The so-called Coulomb-Hankel functions are complex linear combinations of these  solutions:
\begin{eqnarray}
H_L^+(\eta, \varrho) = G_L(\eta, \varrho) + i F_L(\eta, \varrho),\\H_L^-(\eta, \varrho) = G_L(\eta, \varrho) - i F_L(\eta, \varrho)  
\label{eq:coulomb-hankel}
\end{eqnarray}
which describe the incoming ($H_L^-$) and outgoing ($H_L^+$) waves that are being distorted by a Coulomb field.

If a particle is tunneling through a Coulomb barrier, its wave function outside the barrier is exactly described by the outgoing Coulomb-Hankel function, $H_L^+(\eta, \varrho)$. To solve the whole system, one should evaluate the complex internal wave functions based on complex nuclear many-body models and smoothly match them to the exact Coulomb-Hankel functions at the surface of the nucleus \cite{Qi2019}. Within the Gamow model, the internal wave function is mimicked by an effective potential. 

For the decay wave function, the Madelung decomposition is calculated to be
 $$\rho(r) = |H_L^+|^2 = G_L^2(\eta, \varrho) + F_L^2(\eta, \varrho),$$
and
 $$S(r) = \hbar \arctan\left( \frac{F_L(\eta, \varrho)}{G_L(\eta, \varrho)} \right).$$
The Bohmian velocity is
$$v_r(r) = \frac{\hbar k}{m} \left( \frac{1}{G_L^2(\eta, \varrho) + F_L^2(\eta, \varrho)} \right),$$
and the exact quantum potential is given by
\[
\begin{split}
Q(r) = E - \left( \frac{Z_1 Z_2 e^2}{r} + \frac{\hbar^2 L(L+1)}{2m r^2} \right) -\\ -\frac{\hbar^2 k^2}{2m \left[ G_L^2(\eta, \varrho) + F_L^2(\eta, \varrho) \right]^2}.
\end{split}
\]

\subsection{Classical-action reconstruction of the hydrogen problem}

It can be shown that the direct $Q=0$ (classical action) decomposition would fail for the $1/r$ Coulomb potential because the classical action $\phi$ is non-linear in $r$, which forces the density $\rho$ to vary spatially and resurrects the quantum potential ($Q \neq 0$). In addition, for the tunneling problem, there exists a singularity at the classical turning point where the particle's energy equals the potential energy ($E = V_{\text{eff}}$) and the classical momentum $p(r) = 0$.  

Ref.~\cite{LS2026} treats the Coulomb problem (hydrogen atom) by mapping it to a harmonic oscillator via the Kustaanheimo-Stiefel (KS) transformation. Using quaternion coordinates $q = (q_1,q_2,q_3,q_4) \in \mathbb{R}^4$, where $r = q^Tq = |q|^2$, and parameterizing time as $dt' = dt/r$, the Coulomb Hamiltonian transforms to a four-dimensional harmonic oscillator. For bound states ($E < 0$), setting $E = -|E|$ yields:
\begin{equation}
  \frac{p_q^2}{8m} + |E|q^2 = G,
  \label{eq:KS_bound}
\end{equation}
where $G = Ze^2$.
The decomposition is done by integrating over all initial conditions $q_o$:
\begin{equation}
  \psi(q,t') = \int \sqrt{\rho(q,t'|q_o)}\,e^{i\phi(q,t'|q_o)/\hbar} \cdot \psi_o(q_o)\,dq_o.
\end{equation}
The action $\phi$ is explicitly quadratic in $q$ and initial coordinates $q_o$:
\begin{eqnarray}
  \phi_j = 4m\omega^2\!\left[
    \frac{(q^Tq + q_o^Tq_o)\cot(\omega t')}{2}
    - \frac{q^Tq_o}{\sin(\omega t')}
  \right]\nonumber \\ + \frac{4m\omega^2}{2}t' + Gt'',
  \label{eq:phi_coulomb}
\end{eqnarray}
where $\omega = G/\hbar k$ and $t'' = t' + 2\pi k/\omega$. 

The classical density is computed from the exact propagator of the four-dimensional harmonic oscillator. Consequently, the classical amplitude $\sqrt{\rho}$ at fixed $t'$ is a Gaussian in $q$:
\begin{equation}
  \sqrt{\rho}(q,t'|q_o) \propto \exp\!\left(-\alpha(t') q^2 + \beta(q_o,t') q\right).
  \label{eq:rho_Gauss}
\end{equation}

The quantum potential for an individual branch is strictly non-zero. Computing $Q$ for the Gaussian density~\eqref{eq:rho_Gauss} yields:
\begin{equation}
  Q = -\frac{\hbar^2}{2m}\frac{\partial_q^2\sqrt{\rho_j}}{\sqrt{\rho_j}} 
      = -\frac{\hbar^2}{2m}\bigl[(-2\alpha q + \beta)^2 - 2\alpha\bigr].
  \label{eq:Q_Gauss_eval}
\end{equation}
It is a quadratic function of $q$, meaning individual classical trajectories do not satisfy the exact Schr\"{o}dinger equation. However, the non-zero $Q_j$ contributions from individual branches interfere destructively, recovering the correct Bohmian $Q$ for the global wave function.

For the ground state of the 4D harmonic oscillator, the integrated density is a pure Gaussian centered at the origin, $\psi \propto \exp(-\alpha q^2)$. Transforming back to 3D Cartesian coordinates using $r = q^2$, this wave function becomes exactly:
\begin{equation}
  \psi_{0s}(r) \propto e^{-\alpha r},
\end{equation}
which perfectly replicates the exponential decay of the exact Hydrogen $1s$ (or $0s$ in modified notation) wave function, $e^{-r/a_0}$.

\subsection{Tunneling through a repulsive Coulomb potential}

We now examine whether this KS transformation can resolve the unbound state of a positive-energy particle in a pure repulsive Coulomb potential ($V(r) = +G/r$).

For a state with $E > 0$, the classical Hamiltonian with the KS transformation ($r = q^Tq$) becomes:
\begin{equation}
  \frac{p_q^2}{8m} - E q^2 = -G.
  \label{eq:KS_unbound}
\end{equation}
Because $E > 0$, the quadratic potential term is negative. The repulsive Coulomb potential maps exactly to a 4D \emph{inverted} harmonic oscillator. 

The decomposition is analogously performed by integrating over all initial conditions $q_o$:
\begin{equation}
  \psi(q,t') = \int \sqrt{\rho(q,t'|q_o)}\,e^{i\phi(q,t'|q_o)/\hbar} \cdot \psi_o(q_o)\,dq_o.
\end{equation}
To evaluate the action for the inverted oscillator, the frequency $\omega$ must be analytically continued to an imaginary value, $\omega \to i\Omega$ (where $\Omega \propto \sqrt{E}$).

The classical action $\phi$ becomes complex but remains explicitly quadratic in $q$ and initial coordinates $q_o$:
\begin{eqnarray}
  \phi_j = -i(4m\Omega^2)\!\left[
    \frac{(q^Tq + q_o^Tq_o)\coth(\Omega t')}{2}
    - \frac{q^Tq_o}{\sinh(\Omega t')}
  \right] \nonumber\\
  - \frac{4m\Omega^2}{2}t' + Gt''.~~~~
  \label{eq:phi_coulomb_repulsive}
\end{eqnarray}
The classical density is computed from the exact propagator of the inverted four-dimensional harmonic oscillator. Because the action is complex, the exponential arguments shift, but the classical amplitude $\sqrt{\rho}$ at fixed $t'$ remains a Gaussian-like exponential in $q$:
\begin{equation}
  \sqrt{\rho}(q,t'|q_o) \propto \exp\!\left(-\gamma(t') q^2 + \delta(q_o,t') q\right),
  \label{eq:rho_Gauss_repulsive}
\end{equation}
where $\gamma$ and $\delta$ are now complex coefficients derived from the hyperbolic evolution.

As with the bound state, the quantum potential for an individual classical branch is strictly non-zero. Computing $Q$ for the density~\eqref{eq:rho_Gauss_repulsive} yields:
\begin{equation}
  Q = -\frac{\hbar^2}{2m}\frac{\partial_q^2\sqrt{\rho_j}}{\sqrt{\rho_j}} 
      = -\frac{\hbar^2}{2m}\bigl[(-2\gamma q + \delta)^2 - 2\gamma\bigr].
  \label{eq:Q_Gauss_eval_repulsive}
\end{equation}
Because $Q$ is a quadratic function of $q$, individual classical trajectories of the inverted oscillator do not satisfy the exact Schr\"{o}dinger equation. 
However, because the complex action~\eqref{eq:phi_coulomb_repulsive} is strictly quadratic in both $q$ and $q_o$, the integral over $q_o$ remains an exact Gaussian integral. The non-zero $Q_j$ contributions from individual branches interfere destructively. 

Transforming the integrated wave function back to 3D Cartesian coordinates ($r = q^2$) perfectly recovers the exact analytical Coulomb scattering wave function (the confluent hypergeometric function), demonstrating that the global wave function satisfies the correct Bohmian $Q$ despite the failure of $Q=0$ on local branches.

\subsection{Coulomb barrier penetration in nuclear fusion}

We have previously discussed alpha decay, where the external wave is represented by an outgoing Coulomb-Hankel function. The inverse physical process of nuclear fusion into a bound compound nucleus can be described by the same type of Coulomb barrier but with an absorption at the nuclear surface $R$. For two nuclei with reduced mass $\mu$, charges $Z_1e$ and $Z_2e$ and relative coordinate $r$, the exterior effective potential is given for $r>R$ by
\begin{equation}
    V_{\rm eff}(r)=\frac{Z_1Z_2e^2}{r}+\frac{\hbar^2L(L+1)}{2\mu r^2}.
\end{equation}
The radial part of the Schr\"odinger equation becomes then identical to Equation \ref{eq:schrodinger_reduced}, with a slight adaptation of the parameters, $\varrho=kr$ with $k=\frac{\sqrt{2\mu E}}{\hbar}$ and the Sommerfeld parameter becomes $\eta=\frac{Z_1Z_2c^2}{\hbar}\sqrt{\frac{\mu}{2E}}>0$. 

Let's define the Coulomb-Hankel functions as in Equation \ref{eq:coulomb-hankel}, where $H_L^-$ carries inward radial flux and $H_L^+$ carries outward radial flux. In the relative coordinate, there is a single incoming wave, partially reflected by the barrier and partially absorbed into the compound nucleus, and as such, the exterior fusion wave can be written as
\begin{equation}
    u_L(r)=H_L^-(\eta,\varrho)-S_LH_L^+(\eta,\varrho),
\end{equation}
with $S_L$ a non-unitary elastic S-matrix with $|S_L|<1$, which encodes the absorption at $r=R$. The fusion probability is then defined as $T_L=1-|S_L|^2$.

The Madelung decomposition for this case is given by
\[
\rho(r)=\left|H_L^- -S_LH_L^+\right|^2,
\]
and
\[
S(r)=-\frac{\hbar k(1-|S_L|^2)}{\rho(r)}.
\]
This will yield a quantum potential given by
\[
Q(r)=E-\left[\frac{Z_1Z_2e^2}{r}+\frac{\hbar^2L(L+1)}{2\mu r^2}
\right]-\frac{\hbar^2k^2(1-|S_L|^2)^2}{2\mu\left|H_L^- -S_LH_L^+\right|^4}.
\]
This form is analogous to the decay case, but with the caveat that both incoming and outgoing Coulomb-Hankel functions are considered. A notable limit is the complete-absorption $S_L=0$, where the quantum potential assumes the exact same form as the decay one
\[
\begin{split}
Q(r)=E-\left[\frac{Z_1Z_2e^2}{r}+\frac{\hbar^2L(L+1)}{2\mu r^2}
\right]-\\-\frac{\hbar^2k^2}{2\mu\left[G_L^2(\eta,\varrho)+F_L^2(\eta,\varrho)\right]^2},
\end{split}
\]
up to a substitution $m\rightarrow\mu$.

We now note that in the forbidden region $E<V_{\text{eff}}$, the assumption that $Q=0$ would require that $[S(r)]^2=2\mu(E-V_{\rm eff})<0$, which is not possible for real S. Fusion through the Coulomb barrier therefore has the same consequence as the previously discussed decay. It cannot be reconstructed from real classical Hamilton-Jacobi branches alone, and either the action becomes complex, or the quantum potential must be considered.

\section{Berry Phase}

Another point that is worth discussing is the phenomenon of Berry phase \cite{Berry1984}, which yields a further obstruction to the classical reconstruction, this time from geometric phases. Let 
\[
\hat H(R)|n(R)\rangle=E_n(R)|n(R)\rangle,
\]
with slowly varying parameters $R(t)$. In the adiabatic limit, $|\psi(t)\rangle=e^{i\alpha_n(t)}|n(R(t))\rangle$, which upon substitution into the Schrödinger equation yields
\[
\dot\alpha_n=-\frac{E_n}{\hbar}+i\langle n|\dot n\rangle.
\]
Therefore, a closed circuit $C$ in parameter space produces the Berry phase
\begin{equation}
    \gamma_n[C]=i\oint_C\langle n(R)|\nabla_R n(R)\rangle\cdot dR.
\end{equation}
This phase is a holonomy of the Berry connection $\mathcal A_n(R)=i\langle n(R)|\nabla_R n(R)\rangle$. If the Berry curvature $\Omega_n=\nabla_R\times\mathcal{A}_n$ is nonzero, then $\mathcal{A}_n$ cannot be written globally as the gradient of a scalar action, and the Berry phase cannot be generated by a globally defined Hamilton-Jacobi action branch. Note that the Hamilton-Jacobi action branch $\phi_j$ contributes a phase $e^{i\phi_j/\hbar}$. If $\phi_j$ is a globally defined scalar action on parameter space, then for a closed cycle $\Delta\phi_j=0$, or if it is multi-valued, one may impose $\Delta\phi_j=2\pi\hbar k$, but only integer multiples of $2\pi$ can be retrieved, which are physically trivial. For a spin-$1/2$ system, for example, $\gamma_\pm[C]=\mp\Omega[C]/2$, with $\Omega[C]$ the solid angle swept on the Bloch sphere \cite{Berry1984}. This phase is, in general, not an integer multiple of $2\pi$, which yields a further obstruction to a classical reconstruction.

One could, in principle, reproduce the Berry phase by adding the integral of $\mathcal{A}_n$ on the closed circuit $C$ to the action, but thid creates an incompatibility, since $\mathcal{A}_n$ is defined from the quantum eigenstates themselves and not from a classical Lagrangian trajectory. 

\section{Flux Quantization and Josephson Effect in Superconductors}

A final consideration that illustrates the limits of the classical reconstruction is the one for the case of interference in a dc superconducting quantum interference device (SQUID). This is a result that follows from a series of genuinely quantum phenomena, namely flux quantization, quantum tunneling across a barrier, and global superconducting phase quantization, which are all genuinely quantum effects that the classical action description is not able to account for as is. 

For a superconductor, define the wavefunction $\Psi_s=\sqrt{n_s}e^{i\theta}$, with Cooper pair charge $q=2e$ and mass $M^*=2m_e$. The superfluid velocity is given by
\begin{equation}
    M^*\mathbf v_s=\hbar\nabla\theta-qA.
\end{equation}
The requirement of single-valuedness of $\Psi_s$ gives the quantization
\[
\oint\nabla\theta\cdot d\mathbf l=2\pi n,
\]
and
\[
\Phi+\frac{M^*}{n_sq^2}\oint\mathbf j_s\cdot d\mathbf l=n\frac{h}{q}.
\]
In the bulk of a superconducting ring, the supercurrent $\mathbf j_s\simeq0$, so we arrive at the quantization of the magnetic flux in a superconducting ring, $\Phi=n\frac{h}{2e}\equiv n\Phi_0$.

A classical Hamilton-Jacobi action gives instead the relation
\begin{equation}
    M^*\mathbf v=\nabla\phi-q\mathbf A,
\end{equation}
which for $\mathbf v=0$ gives $\oint\nabla\phi\cdot d\mathbf{l}=q\Phi$. If $\phi$ is single-valued, then the left side vanishes and only $\Phi=0$ follows. If $\phi$ is allowed to be arbitrarily multivalued, then $\Phi$ is continuous. The flux quantization follows only from the quantum single-valuedness of $e^{i\phi/\hbar}$, namely $\Delta\phi=nh$. This leads to the conclusion that flux quantization is not produced by local classical action transport. It is a global quantum phase constraint. Furthermore, in the superconducting bulk, $n_s$ is constant, and so the quantum potential is
\begin{equation}
    Q=-\frac{\hbar^2}{2M^*}\frac{\nabla^2\sqrt{n_s}}{\sqrt{n_s}}=0,
\end{equation}
meaning the effect cannot be attributed to a local quantum potential either.

We now move on to the discussion of the Josephson effect. Let us consider a junction made out of two superconducting electrodes separated by an insulating thin barrier of thickness $d$, which we model as a rectangular barrier. For two superconductors left ($L$) and right ($R$) separated by an insulating barrier, the gauge-invariant phase difference is given by
\[
\delta=\theta_R-\theta_L-\frac{2e}{\hbar}\int_L^R\mathbf{A}\cdot d\mathbf{l}.
\]
Inside a rectangular barrier, the wavefunction can be modeled as
\begin{equation}
    \Psi(x)=a e^{-\kappa x}e^{i\theta_L}+b e^{-\kappa(d-x)}e^{i\theta_R},
\end{equation}
where $\kappa=\frac{\sqrt{2M^*(U_0-E)}}{\hbar}$. The current density is given by
\begin{equation}
    j=\frac{\hbar}{M^*}\operatorname{Im}\left(\Psi_B^*\frac{d\Psi_B}{dx}\right),
\end{equation}
where direct substitution yields
\[
j=\frac{2\hbar\kappa}{M^*}abe^{-\kappa d}\sin\delta,
\]
and defining the critical current density as $j_c=\frac{2\hbar\kappa}{M^*}ab e^{-\kappa d}$, we retrieve the relation $I=I_c\sin\delta$. The exponential dependence $I_c\propto e^{-\kappa d}$ is the tunneling amplitude through the insulating barrier. A real classical Hamilton-Jacobi action would require
\begin{equation}
    (\partial_x\phi)^2=2M^*(E-U_0)<0
\end{equation}
inside the barrier, which is impossible. Therefore, the Josephson effect inherits the same obstruction as ordinary finite-barrier tunneling.

We can finally consider a dc SQUID with two identical junctions $I=I_c\sin\delta_1+I_c\sin\delta_2$, for which flux quantization imposes that $\delta_2-\delta_1=\frac{2\pi\Phi}{\Phi_0}$, leading to
\begin{equation}
    I_c^{\text{SQUID}}(\Phi)=2I_c\left|\cos\left(\frac{\pi\Phi}{\Phi_0}\right)\right|,
\end{equation}
the well-known critical current modulation, a quantum interference effect \cite{PhysRevLett.12.159}. Since this derivation combines the two previously discussed structures of coherent pair tunneling and global superconducting phase quantization, its description is inaccessible through a purely classical-action reconstruction. The effect can only be obtained from local classical branch transport if the quantum phase constraint and tunneling amplitude are inserted a posteriori, which contradicts the first-principles approach that is the stated objective of \cite{LS2026}.

\section{Summary and discussion}
\label{sec:discussion}
%In summary, we have examined the extent to which exact quantum tunneling can be reconstructed from superpositions of classical action branches with associated classical densities, as proposed in Ref.~\cite{LS2026}. While the formalism remains consistent for systems admitting globally defined real multi-valued classical actions,  finite-barrier tunneling introduces essential structures that lie beyond such a description.

In summary, we have examined the extent to which quantum phenomena can be reconstructed from superpositions of classical action branches with associated classical densities. While the formalism remains consistent for systems admitting globally defined real multi-valued classical actions, finite-barrier tunneling and global quantum phase phenomena introduce essential structures that lie beyond such a description.

Exact quantum tunneling through finite barriers appears to require structures that cannot, in general, be reduced to purely real classical action satisfying local Hamilton-Jacobi dynamics. Consequently, both classical actions and probability densities must inevitably admit complex-valued solutions. The restriction of the LS framework to problems admitting globally defined real classical branches also has a direct kinematic consequence inside the barrier. 

One crucial distinction between the semi-infinite potential step and genuine finite-barrier tunneling is the necessary presence of the growing exponential component inside the forbidden region. Unlike the decaying branch, this contribution is fixed globally by the exit boundary condition and encodes nonlocal information about the barrier geometry. Consequently, the exact tunneling solution cannot generally be reconstructed from purely local transport of real classical branches alone. The physically essential ingredient, the growing exponential $Be^{+\kappa x}$ with amplitude set by the exit boundary at $x = a$, is invisible to any construction based on local classical flux conservation at the entry boundary. It is precisely this nonlocal, globally determined component that the quantum potential $Q_{\mathrm{II}}$ encodes. $Q_{\mathrm{II}}$ nearly cancels $V_0 - E$ inside the barrier, flattening the effective Bohmian potential and enabling the guided particle to traverse the barrier in finite time with exponentially small but nonzero probability. The LS construction, which sets $Q = 0$ by ansatz, has no mechanism to represent this effect.

The additional examples considered here show that the obstruction is not limited to tunneling. Berry phase is a holonomy in parameter space and cannot be generated by a globally defined scalar Hamilton–Jacobi action. Flux quantization follows from the single-valuedness of a macroscopic superconducting order parameter, even in regions where the local quantum potential vanishes. Josephson tunneling and dc SQUID interference combine coherent barrier penetration with global superconducting phase constraints. These effects can be reproduced only if the relevant quantum phase information is inserted into the classical-action construction a posteriori. This illustrates a main limitation of the framework, which is that it may reproduce quantum wave functions in special cases where the required phase and amplitude structure is already encoded in the chosen branches, but it does not derive the genuinely quantum structures responsible for tunneling, geometric phase, and superconducting interference, among many other genuinely quantum phenomena, from real classical action alone.

\bibliography{ref}
\end{document}